\documentstyle[prl,aps,twocolumn,epsf]{revtex}
\begin{document}
\draft
\title{
Renormalization group analysis of electrons near a Van Hove singularity}
\author{J. Gonz\'alez $^1$, F. Guinea $^2$ and M. A. H. Vozmediano $^3$ \\}
\address{
	$^1$Instituto de Estructura de la Materia. 
	Consejo Superior de Investigaciones Cient{\'\i}ficas. 
	Serrano 123, 28006 Madrid. Spain. \\
	$^2$Instituto de Ciencia de Materiales. 
	Consejo Superior de Investigaciones Cient{\'\i}ficas. 
	Cantoblanco. 28049 Madrid. Spain. \\
 	$^3$Escuela Polit\'ecnica Superior. 
	Universidad Carlos III. 
	Butarque 15.
	Legan\'es. 28913 Madrid. Spain.}
\date{\today}
\maketitle
\begin{abstract}
\widetext
	A model of interacting two dimensional electrons near a Van Hove
singularity is studied, using renormalization group techniques.
In hole doped systems, the chemical potential is found to be pinned
near the singularity, when the electron-electron interactions
are repulsive. The RG treatment of the  
leading divergences appearing in perturbation theory
give rise to marginal behavior
and anisotropic superconductivity.
\end{abstract}
\narrowtext
	The relevance of a Van Hove singularity for the high-T$_c$
compounds was pointed out shortly after 
their discovery\cite{LB,Schultz,LR,Friedel,Dzy},
and a more refined ^^ ^^ Van Hove scenario " 
was proposed afterwards\cite{Newns}.
In addition there is mounting numerical evidence that such a
singularity arises in studies of strongly 
correlated 2D systems\cite{BSW,PHL,DNM}.
This fact is in agreement with photoemission studies in hole doped
high-T$_c$ materials\cite{photo}.

	The dynamics of interacting two dimensional electrons 
near a van Hove singularity are anomalous, as a standard perturbative
treatment is plagued by logarithmic divergences. The study of these
divergences requires a renormalization group treatment. The feasibility
of such an approach in condensed matter systems has been analyzed in 
ref.\cite{Shankar}. In the following, we apply a suitably generalized
version of the scheme proposed in\cite{Shankar} to the problem at hand.

\begin{figure}
\epsfysize=7cm\epsfbox{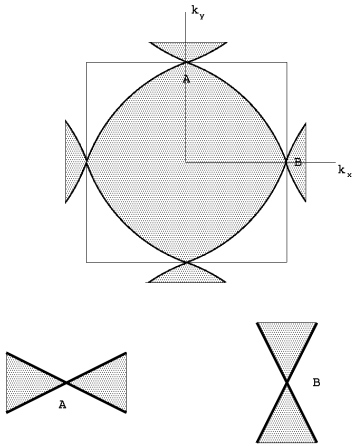}
\caption{Schematic picture of the Brillouin Zone and the Fermi
surface of the system considered in the text}
\label{BZ}
\end{figure}
	A sketch of the band structure to be discussed is shown 
\break
\newline\bigskip\bigskip\bigskip\bigskip\bigskip\bigskip
\noindent

in fig. \ref{BZ}.

	In a square lattice there are two inequivalent points in
the Brillouin Zone where a van Hove singularity is likely to occur.
Near these singularities, the dispersion relation is quadratic,
and the Fermi surface is well approximated by two intersecting 
straight lines. We will not consider the case of perfect nesting, when
the Fermi surface is straight throughout the entire Brillouin
Zone.
	If we restrict our attention to the electronic states near these
two singularities, we can write a long wavelength, effective hamiltonian as:

\begin{eqnarray}
	{\cal H} &= &\sum_{i,j = A,B} - \int  
	\Psi_i^{\dag} ( \vec{r} ) \left[ 
	\frac{\hbar^2}{2 m_{x,i}} \partial_x^2 -
	\frac{\hbar^2}{2 m_{y,i}} \partial_y^2 \right]
	\Psi_i ( \vec{r} ) dx dy \nonumber \\ &+
	&u_{i,i';j,j'} \int   
	\Psi_i^{\dag} ( \vec{r} ) \Psi_{i'}^{\dag}
	( \vec{r} ) \Psi_j ( \vec{r} ) \Psi_{j'} ( \vec{r} ) dx dy
\label{hamil}
\end{eqnarray}

	where $m_{x,A} = m_{y,B}$ and $m_{y,A} = m_{x,B}$ 
(see figure \ref{BZ}).
In the following, we will use the parameters
$\overline{m} = ( m_{x,A} + m_{y,A} ) / 2$ and $\Delta m = 
( m_{x,A} - m_{y,A} ) / 2$.
We have omitted spin indices for simplicity. 
The number of possible 
(isotropic) interactions is restricted by symmetry. An
umklapp interaction of the type
$u_{A \uparrow , A \downarrow ; B \uparrow , B \downarrow}$
is allowed. A term like this transfers a pair of electrons from the
vicinity of one of the singularities to the vicinity of the other.
The total momentum transfer coincides with a lattice reciprocal 
vector. 

	The couplings in (\ref{hamil})are 
expressed in units of [ energy $\times$ area ].
For instance, for a local Hubbard-like interaction, $U$, the parameters
in (\ref{hamil}) are proportional to $U \times$
the area of the unit cell. By scaling units
in (\ref{hamil}), all couplings can be made dimensionless,
that is, $\tilde{u} = \overline{m} u / \hbar^2$. When
(\ref{hamil}) is a low energy approximation to a Hubbard like
hamiltonian, these dimensionless  parameters are proportional 
to $\frac{U}{W}$, where $U$ is the Hubbard interaction and $W$ is
the bandwidth.

	We first consider the electrons near a given van Hove singularity.
An anisotropic scaling of lengths
leads to $\Delta m = 0$. 
The fact that the interaction is dimensionless
implies that perturbative corrections should depend logarithmically on the
remaining scales implicit in (\ref{hamil}): the high energy cutoff, and the
low energy scale, that is, the temperature or the external frequency in the 
diagram.

\begin{figure}
\epsfysize=7cm\epsfbox{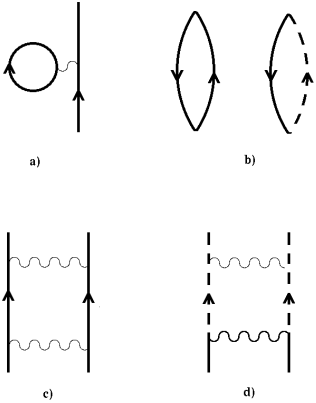}
\caption{Diagrams calculated in the text. Full and broken lines
correspond to propagators of electrons near the two inequivalent
singularities. Thin and thick wavy lines denote intra- and 
inter-singularity scattering.}
\label{diagram}
\end{figure}

	For instance, the particle 
hole polarizability is shown in fig. (\ref{diagram}b). Using a cutoff in energies,
such that the filled states in the Fermi sea have energies
$- \epsilon_c < \epsilon < 0$, we obtain:

\begin{eqnarray}
	\hbox{Re} \chi ( \vec{q} , \omega ) &= 
	&- \frac{1}{2 \pi^2}  \left[ \log \left| \frac{
	4 \epsilon_c \epsilon_q }
	{  \omega^2 - \epsilon_q^2 } \right|  
	- \frac{\omega}{\epsilon_q}
	\log \left| \frac{\omega + \epsilon_q}{\omega - \epsilon_q}
	\right|  + 2 \right] \nonumber \\
	\hbox{Im} \chi ( \vec{q} , \omega ) &= 
	&\frac{2}{\pi \epsilon_q} \left( | \omega + \epsilon_q | -
	| \omega - \epsilon_q | \right)   
\label{susc}  \\ & &
	\; \; \; \; \; \; \; \; \; \; \; \; \; \; \; 
	\; \; \; \; \; \; \; \; \; \; \; \; \; \; \; 
	\; \; \; \; \; \; \; \; \; \; \; \; \; \; \; 
	\omega , \epsilon_q \ll 
	\epsilon_c \nonumber
\end{eqnarray}
	where $\epsilon_q = \frac{\hbar^2 ( q_x^2 - q_y^2 )}{2 \overline{m}}$
(note that $\vec{q}$ is measured from the singularity).
The angular dependence on $q$ is determined by the SO(1,1) symmetry of the
dispersion relation. Other choices of cutoff, which break this symmetry,
do not change the low energy behavior of the polarizability,
given by (\ref{susc}). This expression agrees with previous estimates
of the polarizability of electrons near a van Hove 
singularity\cite{Newns1,GGA,HR}. The imaginary part of the susceptibility
is consistent with the ^^ ^^ marginal Fermi liquid " hypothesis\cite{MFL}.

	The inter-singularity polarizability (also shown 
in fig. \ref{diagram}b)
diverges as:

\begin{eqnarray}
	\hbox{Re} \chi ' ( \vec{q} , \omega ) &\sim &\log \left(
	\frac{\overline{m}}{\Delta m} \right) \log \left[
	\frac{\epsilon_c}{\hbox{max} 
	\left( \omega , \frac{\hbar^2 ( q_x^2 + q_y^2 )}{ 2 \overline{m}
	}\right)} \right] \nonumber \\
	\hbox{Im} \chi' ( \vec{q} , \omega ) &\sim 
	&\log \left( \frac{\overline{m}}{\Delta m} \right)
	\hbox{min} \left[ 1 , \frac{\overline{m} \omega}{\hbar^2 ( q_x^2
	+ q_y^2 )} \right] \label{suscp}  \\ & &
	\; \; \; \; \; \; \; \; \; \; \; \; \; \; \; 
	\; \; \; \; \; \; \; \; \; \; \; \; \; \; \; 
	\; \; \; \; \; \; \; \; \; \; \; \; \; \; \; 
	\omega , \epsilon_q \ll 
	\epsilon_c \nonumber
\end{eqnarray}

	The electron-hole polarizabilities screen the bare interactions.
This flow of the effective couplings makes the problem similar to the
1D Luttinger liquid\cite{graphite}. 
An unrelated renormalization scheme was
proposed in\cite{ZB}, on the basis of the 
marginal Fermi liquid phenomenology\cite{MFL}. 
The $q$ dependence of the diagrams derived
from (\ref{hamil}), and the electron-electron scattering 
channels are, however, totally different from those proposed
in\cite{ZB}.

	The intra-singularity
effective interaction, to lowest order, is:

\begin{equation}
	u_{eff} ( q ) \approx \frac{u_0}{1 + \frac{\tilde{u}}{\pi^2}
	\log \left( \frac{\epsilon_c}{\epsilon_q} \right) }
\label{ueff}
\end{equation}

	This result implies that the system is unstable against any
distortion with a (small) wavevector along the Fermi line. Repulsive
interactions favor a spin density wave, while attractive interactions
lead to a charge density wave. Using the same argument for the
effective inter-singularity interactions, screened by $\chi'$, we 
also find instabilities for the wavevector $( \pi , \pi )$ which
connects the two special points.

	To lowest order, we need also consider the renormalization of
the self-energy, given by the Hartree diagram shown in fig. (\ref{diagram}a).
In a conventional Fermi liquid, the integrating out of the
states in a given energy shell leads to a constant contribution,
which uniformly shifts the one
electron levels. The present case differs, in that
each new energy shell contains more and more states, and the
shift increases as the renormalization proceeds towards 
the van Hove singularity. For instance, let us define the Fermi
energy, $\epsilon_F$, referred to the singularity.
In dimensionless units, the flow of $\tilde{\epsilon}_F =
\epsilon_F / \epsilon_c$ is given by:

\begin{equation}
	\frac{\partial \tilde{\epsilon}_F}{\partial l} =
	\tilde{\epsilon}_F - \tilde{u} \log \left(
	\frac{1}{1 - \tilde{\epsilon}_F} \right)
\label{Fermi}
\end{equation}

	where $l = - \log ( \epsilon_c )$.

\begin{figure}
\epsfysize=9cm
\epsfbox{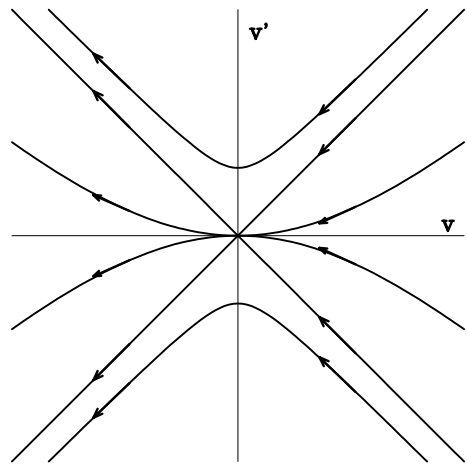}
\caption{Schematic view of the flow of the intra- (v) and 
inter-singularity scattering (v') in the BCS channel.}
\label{flow}
\end{figure}

	The first term in(\ref{Fermi}) is determined by dimensional
arguments. The Fermi energy seems larger when measured in terms
of the reduced cutoff. The second term gives the shift due to the
interaction with the high energy part of the Fermi sea.We assume that
the high energy states which are removed lie at energy $\epsilon_c$
below $\epsilon_F$. 
Unlike in 
normal metals, the second term does not scale in step with 
the first\cite{Shankar}. Thus, for repulsive interactions and
$\epsilon_F > 0$, there may be a non trivial, {\it stable}, fixed 
point, at which the
flow in (\ref{flow}) is arrested. As $\epsilon_c$ is further reduced,
$\tilde{\epsilon}_F = {\tilde{\epsilon}_F}^*$ remains unchanged.
In dimensionful units, it means that the Fermi energy gets pinned
at the van Hove singularity.

\begin{figure}
\epsfysize=7cm\epsfbox{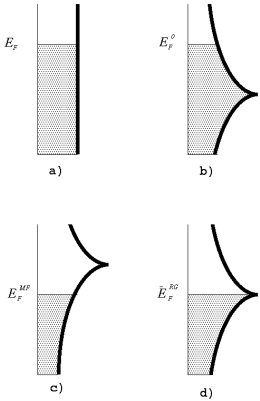}
\caption{Schematic view of the Fermi level pinning process
discussed in the text. a) Density of states in the dopant planes.
b) Density of states in the CuO planes in the absence of the
electron-electron repulsion. c) As in b), with the interaction treated 
within mean field. d) As in b), with the interaction treated within 
the RG scheme presented in the text.}
\label{doping}
\end{figure}

	A sketch of the physics involved is shown in fig. (\ref{doping}).
The chemical potential at the CuO planes is fixed by the other
planes (fig. \ref{doping} a). We know that, for hole doped systems,
a mean field treatment of the interaction places the singularity
above the Fermi level (fig. \ref{doping} c). It implies that, in the
absence of the (repulsive) interaction, the singularity should be
at lower energies ( fig. \ref{doping} b). Let us assume that the
singularity is below the Fermi level when the interaction is switched off.
Then, if the Hartree diagram is treated using renormalization group
techniques, 
starting from the previous situation, the Fermi
energy will get pinned at the singularity (fig. \ref{doping} d).

	The difference between a mean field and a RG treatment of
the Hartree potential lies in the fact that, in the latter case, electrons
near the Fermi energy experience a larger shift from ^^ ^^ fast " electrons,
which lie at high energies, than from ^^ ^^ slow " electrons. The RG
analysis allows for the correlation hole generated by the low energy
electrons. The pinning discussed here is only possible in
nominally hole doped systems, in agreement with photoemission
experiments\cite{photo}.

	In the absence of a BCS instability (to be discussed below),
the low order processes that we have considered so far renormalize down the
bare interactions, in the way shown in (\ref{ueff}). The hamiltonian
(\ref{hamil}) flows towards a free fixed point. In the renormalization
process, however, other quantities, such as the quasiparticle weight,
are also modified (wavefunction renormalization). Because of the
logarithmic divergences in the polarizabilities, the quasiparticles
are renormalized 
throughout the Fermi surface. The general form of this renormalization
is: 

\begin{equation}
	1 - Z_k = \frac{\partial \hbox{Re} 
	\Sigma ( k , \omega )}{\partial \omega}
	\sim {\tilde{u}}^2 \log \left( 
	\frac{\epsilon_c}{\epsilon_k} \right)
\label{wvren}
\end{equation}

	Thus, the spectral weight of the quasiparticle states is
energy dependent, and decreases as the Fermi energy is approached.
This behavior is reminiscent of a 1D Luttinger liquid\cite{graphite},
and qualitatively agrees with the ^^ ^^ marginal Ferm liquid "
hypothesis\cite{MFL}. In addition, the transport properties, in the normal
state, of the electrons near a singularity are strongly modified
by the existence of a Umklapp electron electron 
scattering channel\cite{LR,umklapp}.

	Finally, we analyze the BCS channels, shown in 
figure (\ref{diagram}c) and (\ref{diagram}d).
These diagrams also induce logarithmic corrections to the couplings,
which diverge like
the logarithm of the cutoff squared. Thus, this effect dominates over the
influence of the electron-hole renormalization from the
diagrams in fig. (\ref{diagram}b).

	Because of this anomalous divergence, a straightforward approach 
of RG techniques is not possible. We have chosen to scale the couplings 
as function of $l^2 = \log^2
( \epsilon_c )$ which corresponds to summing the 
leading $\log^2$ divergences\cite{Schultz}. 
Defining $v$ as the intra-singularity scattering,
and $v'$ as the inter-singularity scattering, the flow
is given by:

\begin{eqnarray}
	\frac{\partial v}{\partial l^2} &= &- v^2 - {v'}^2 \nonumber \\
	\frac{\partial v'}{\partial l^2} &= &- 2 v v'
\label{flowc}
\end{eqnarray}
 
	These equations can be integrated, and the flow follows
the lines $v' = \pm \left( k \pm \sqrt{ k^2 + v^2} \right)$, where $k$ is
a constant. The flow of the couplings is schematically 
shown in fig. (\ref{flow}).

A purely repulsive $v$ is renormalized towards $v = 0$, which
is a marginal fixed point. If $| v'| > | v |$, the couplings eventually become
large and attractive. The flow towards strong coupling
is the renormalization group version of the BCS instability.
Thus, this instability always takes place when $| v'| > | v |$, irrespective
of the sign of the couplings. 

	The instability arises because a large $v'$ favors a
coherent superposition of Cooper pairs at both singularities.
A repulsive interaction ( $v' > 0$ ) implies that the condensate
amplitude has opposite signs at the two singularities, i. e.
d-wave pairing.

\begin{figure}
\epsfxsize=7cm\epsfbox{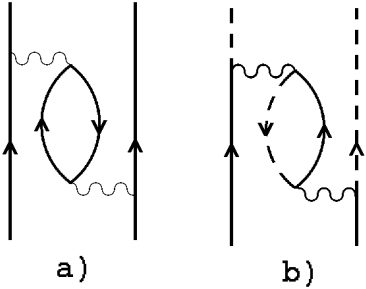}
\caption{Kohn-Luttinger diagrams which renormalize
the couplings in the BCS channel.}
\label{Kohn}
\end{figure}

	Usually, a repulsive interaction in $q$ space is, either constant, 
or decays for large $q$'s. In the present case, however, 
$v$ and $v'$ are screened in very different ways by the
electron-hole polarizabilities calculated in (\ref{susc}) and 
(\ref{suscp}). The relevant diagrams are shown in fig. (\ref{Kohn}).
Intra-singularity screening should dominate (provided that
$\log \left( \frac{\overline{m}}{\Delta m} \right) \sim O ( 1 )$),
 as can be appreciated by comparing (\ref{susc}) and (\ref{suscp}).
These processes are the equivalent to those
originally discussed by Kohn and Luttinger\cite{KL}, when studying
the possibility of superconductivity from screened repulsive
interactions. The physics is very much the same,
except that, in the present case,
the anisotropy of the screening matches
well the phase space available for the formation of Cooper
pairs. Thus, the instability arises in a more natural way.
In addition, the existence of a van Hove singularity modifies
the relation between the BCS gap and the dimensionless
coupling, which now becomes $\Delta_{BCS} \sim
T_c \sim \epsilon_c e^{- 1 / \sqrt{ v' -  v }}, ( v , v' > 0)$.

	In summary, we have shown that a system of interacting electrons
near a 2D van Hove singularity can be analyzed using renormalization
group methods. The treatment presented here is still far from complete, 
but the derived features are
independent of the details of the model, and consistent with
the observed behavior of high-T$_c$ materials.

\end{document}